\documentclass[%
 reprint,
superscriptaddress,
 amsmath,amssymb,
 aps,
 pra,
]{revtex4-1}

\usepackage{graphicx}% Include figure files
\usepackage{dcolumn}% Align table columns on decimal point
\usepackage{bm}% bold math

\usepackage{amsmath, amssymb, graphics}

\newcommand{\mathsym}[1]{{}}

\usepackage{hyperref}% add hypertext capabilities
\begin{document}

\preprint{APS/SEPS}

\title{Heavily enhanced dynamic Stark shift in a system of Bose-Einstein condensation of photons}
%\thanks{A sample paper.}%

\author{Weikang Fan}
\author{Miao Yin}
 \altaffiliation{Corresponding author; scmyin@scut.edu.cn}
 \affiliation{Department of Physics, South China University of Technology, Guangzhou 510640, People's Republic of China}
\author{Ze Cheng}
 \affiliation{School of Physics, Huazhong University of Science and Technology, Wuhan 430074, People's Republic of China}

\date{\today}% It is always \today, today,
             %  but any date may be explicitly specified

\begin{abstract}
The dynamic Stark shift of a high-lying atom in a system of Bose-Einstein condensation (BEC) of photons inside a two-dimensional microcavity is discussed within the framework of nonrelativistic quantum electrodynamics (QED) theory. It is found that the Stark shift of an atom in BEC of photons is modified by a temperature dependent factor $F_{\text{BEC}}$, compared to that in a normal two-dimensional photonic fluid. In photonic BEC, the value of Stark shift is always greater than that in two-dimensional free space. Physical origin of this phenomenon is presented and potential application is also discussed.
\end{abstract}

% PACS, the Physics and Astronomy
                             % Classification Scheme.
%\keywords{Suggested keywords}%Use showkeys class option if keyword
                              %display desired
\maketitle

%\tableofcontents

\section{Introduction}
%---------------------------------REVIEW ON BEC AND INTRO 2 PHOTONIC ONE------------------
Bose-Einstein condensation (BEC), predicted by Einstein \cite{EA24} in 1924, is a remarkable state of matter. For a Bose system at low temperature, significant fraction of the particles condensates into zero momentum state with minimum free energy of the system.
BEC was first observed in rubidium by Anderson \emph{et al.} \cite{anderson95} in 1995. Thereafter, the phenomenon was revealed in many other systems of atoms or quasi-particles, such as sodium \cite{davis95}, lithium \cite{Sto96}, cesium \cite{weber03}, potassium \cite{modugno01}, hydrogen \cite{fried98}, polariton \cite{amo09}, etc. However, BEC in the simplest Bose system, the photon system, was not observed until 2010 \cite{klaers10}.
What impeded us so long from realizing Bose-Einstein condensate (BEc, lower case ``c'' to distinguish from BEC) of photons is that in a normal blackbody radiation cavity, the photon is massless and the photon number is not conserved which leads to a vanishing chemical potential. In 2010, Klaers \textit{et al.} \cite{klaers10} have overcome both obstacles by confining laser pump light in a two-dimensional microcavity which is filled with dye and bounded by two highly reflective concave mirrors. They established the conditions required for the light to thermally equilibrate as a gas of conserved particles rather than as an ordinary blackbody radiation.
%--------------------------------ATOMIC WORKS AND INSPIRATION---------------------------------
\par As is well known, quantum optical effects of atoms are not only dependent on their internal structures, but also on external electromagnetic environment. People have explored various systems in modified electromagnetic environment such as dielectric medium\cite{MR00,SMSL98}, photonic crystals\cite{JSWJ90,ZSY00}, and optical microwave guides\cite{CSMT01}. The results show that the energy-level shift is modified accordingly. For example, Wang \textit{et al.}\cite{LambPC} predicted that the dominant contribution to the Lamb shift comes from the emission of real photons in photonic crystals and the Lamb shift can be enhanced by 1 or 2 orders of magnitude, termed as `giant' Lamb shift. In a recent work, we investigated the Lamb shift of a hydrogen atom inside a Kerr nonlinear blackbody (KNB) and also found that the modification of the Lamb shift is a `giant' one \cite{YM09L}. What is more, we found that the dynamic Stark shift in KNB reveals a temperature dependency \cite{YM09S}. Inspired by the experimental demonstration of BEC of photons, in this paper, we aim to investigate the dynamic Stark shift of an atom in a system of BEC of photons inside a two-dimensional microcavity. We find that in such a system, the dynamic Stark shift of a high-lying atom can be heavily enhanced.
%--------------------------------NEED APPLICATIONS TO ENRICH----------------------------------
%--------------------------------OUTLINE OF THE WORK------------------------------------------
\par The remainder of this paper is organized as follows. In Sec. II, we configure the microcavity for BEC of photon system and theoretically model the BEC of weakly interacting photons in a two-dimensional optical cavity. The expression of dynamic Stark shift of a high-lying atom in photonic BEc is derived in Sec. III. In Sec. IV, we discuss the modified Stark shift and the role of dimension. Finally, we make a brief conclusion in Sec. V.
\section{Bose-Einstein Condensation of Photon System}
In this section, we theoretically establish the model for BEC of a two-dimensional photon gas. We first base the design on Klaers' work to configure the two-dimensional microcavity in which photonic BEC is attained. Then we turn to reexamine the dispersion relations of two-dimensional free and weakly-interacting photon gas. Finally, we discuss the conditions of BEC of photons.
\subsection{Microcavity Resonator}
As is known to all, a prerequisite of BEC is that the chemical potential should be non-vanishing \cite{BEC03}. However, chemical potential of the photon system is vanishing in normal blackbody radiation, that is, the photon number is not conserved when the temperature of the photon gas is varied. For normal blackbody radiation, photons disappear in the cavity walls instead of occupying the cavity ground state \cite{klaers10}. Luckily, number conserving thermalization of two-dimensional photon gas was experimentally achieved by Klaers \textit{et al.} \cite{klaers10}.
\begin{figure}[htbp]
\small
\centering
\includegraphics[width=3.2 in]{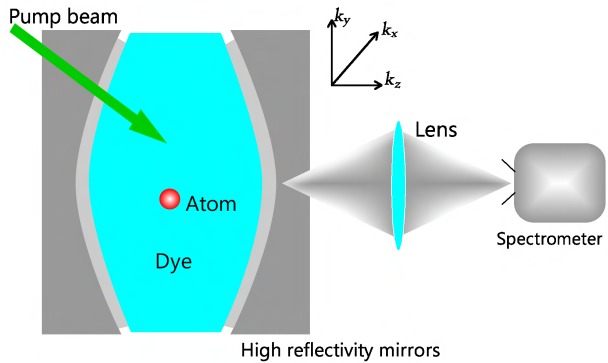}
\caption{(Color online) Schematic of the optical microcavity. The cavity consists of two high-reflectivity curved mirrors which is fulfilled with dye solution. An atom is immersed inside the microcavity and is located away from the laser path.}
\label{fig:Config}
\end{figure}
\par Typically, a microcavity resonator is required to be composed of highly reflective mirrors with reflectivity over $99.997\%$ at both ends, acting as the walls, as shown in FIG. 1. Separation between surfaces of adopted curved mirrors is described by $D(r)= D_0 - 2(R-\sqrt{R^2-r^2})$ with $r$ being the distance from the optical axis, $D_{0}$ being the mirror separation at distance $r = 0$, and $R$ the radius of the curvature. A laser of $532~$nm pumps dye onto the cavity at an angel of $45^{\circ}$ to the optical axis. The atom should be located inside the cavity and completely immersed in the dye solution. Also, it is away from pump beam to avoid possible pump light excitations. Generally, the atom couples with photons to form polaritons, a bosonic quasi-particle.
However, owing to the frequent collisions of photons and dye solution, the coupling is broken\cite{TKYM89,DA00}. Thus we can convince ourselves that the BEC here is real BEc of photons instead of that of polaritons.
\subsection{Free Photon Dispersion Relation}
The mirrors have confined the photons to a two-dimensional plane, simultaneously imposing a boundary condition on both sides of inside electric field. The allowed $z$-component, which is along the optical axis of the wave vector of photon, is $k_z(r)=n'\pi/D(r)$ where $n'$ is an integer.
The electric field vanishes at the reflecting surface thus imposing a quantization condition for $z$-component of the wave vector. Photons are trapped inside the resonator. In this two-dimensional plane, a plane-mode wave propagates at small angle with respect to $z$ axis, the effective mass of a non-relativistic particle is $m_{\gamma}=\hbar\omega/c^{2}$ \cite{Deu91}.
\par For a two-dimensional noninteracting photon gas inside the cavity resonator, the de Broglie dispersion relation or energy-momentum relation under small-angel approximation is given by
\begin{equation} \label{FBHamiltonian}
\epsilon(\bm{p})=m_{\gamma} c^2+\frac{\bm{p}^2}{2m_{\gamma}}.
\end{equation}
We note that in the present cavity resonator $k_{z}$ is greater than $k_x$, $k_y$ in the limit of small-angle propagation. The small transverse momentum of the photon is given by
\begin{equation}
\bm{p}^2=p_x^2+p_y^2.
\end{equation}
Confining our sight to the two-dimensional plane, we can view free photons as normal particles in the transverse plane.
\subsection{Bogoliubov Dispersion Relation for the Photon Gas}
There exists BEC in an ideal Bose system at absolute zero temperature. This feature survives in the case of weakly interacting Bose system. Since as the interaction vanishes, one should recover the BEc state\cite{Chiao00}. From the work by Klaers \textit{et al.}\cite{klaers10}, knowing the intensity of light $I(r)$ and relative refraction index $n_{r}$, we may find the expression of the interaction energy of photons as
\begin{equation}\label{E_int1}
E_{\text{int}}=m_{\gamma}c^2 n_{r} I(r).
\end{equation}
By using the method applied in Gross$-$Pitaevskii theory, we introduce a dimensionless parameter $g=-m_{\gamma}^4c^6/2\pi h^3 \tilde{n} q$ to describe the strength of interaction \cite{Had09}, with $c$ being the speed of light, $\tilde{n}$ the quantum concentration of current system, and $q$ a constant. Subsequently, the interaction energy $E_{\text{int}}$ can be rewritten as
\begin{equation}
E_{\text{int}}=\hbar^2/m_{\gamma} g N_0 \tilde{n}^2.
\end{equation}
In this cavity, $g \sim (7\pm3)\times 10^{-4}$, far below that reported in a normal two-dimensional quantum gas experiment ($-2\sim-1$ orders of magnitude).
This indicates that the photons are weakly interacting which enable us to apply Bogoliubov transform in such a two-dimensional quantum gas system.
\par Since the present system is open, i.e., it is connected to an external reservoir of particles, the average of the total particle number fluctuates around an average value. Thus the total particle number needs only be conserved on the average. To formally illustrate the effect, we introduce the Lagrange multiplier method and
subtract a chemical potential term $\mu N_{\text{open}}$ from the Hamiltonian of the system, as usually done in statistical mechanics,
\begin{equation}\label{InteractionTermCanonical}
H=H' - \mu N_{\text{open}},
\end{equation}
where $N_{\text{open}}=\sum_{\bm{p}}a_{\bm{p}}^{\dag}a_{\bm{p}}$ is the total number operator and $\mu$ the chemical potential. Experimentally, pump beam in the present configuration act as physical implementation of the external reservoir. $H'$ in Eq.\eqref{InteractionTermCanonical} is given by
\begin{equation}\label{PhotonH}
H'=H_{\text{int}}+H_{\text{non-int}}.
\end{equation}
For a free Bose system, the Hamiltonian is
\begin{equation}\label{NIpH}
H_{\text{non-int}}=\sum_{\bm{p}}\epsilon(\bm{p}) a_{\bm{p}}^{\dag} a_{\bm{p}},
\end{equation}
where $a_{\bm{p}}^{\dag}$ and $a_{\bm{p}}$ are creation and annihilation operators with momentum $\bm{p}$, respectively.
We know that in a weakly-interacting Bose system near absolute zero temperature, the particles in zero momentum state are in great number $N_{0}$. For such ground state $\left|N_{0}\right\rangle$, the zero-momentum operator $a_{0}^{\dag}$ and $a_{0}$ obey the following relations
\begin{equation}
\begin{aligned}
a_{0}\left|N_{0}\right\rangle &= \sqrt{N_{0}}\left|N_{0}-1\right\rangle,\\
a_{0}^{\dag}\left|N_{0}\right\rangle &= \sqrt{N_{0}+1}\left|N_{0}+1\right\rangle .
\end{aligned}
\end{equation}
Since $N_{0}$ is large enough in this case, we can approximately take $\sqrt{N_{0}+1}\thickapprox \sqrt{N_{0}}$.
The operators also satisfy Bose commutation relations\\
\begin{equation}\label{BoseCMTRelation}
\begin{aligned}
\left[a_{\bm{p}},a_{\bm{q}}^{\dagger}\right] & = \delta_{\bm{p},\bm{q}},\\
\left[a_{\bm{p}},a_{\bm{q}}\right] & = \left[a_{\bm{p}}^{\dagger},a_{\bm{q}}^{\dagger}\right]=0.
\end{aligned}
\end{equation}
\par The interaction term describes the momentum transfer between the photons arising from the potential energy $V(\bm{\kappa})$. It represents the annihilation of two photons with momenta $\bm{p}$ and $\bm{q}$, along with the creation of two photons with momenta $\bm{p+\kappa}$ and $\bm{q-\kappa}$. The interaction Hamiltonian can then be written as
\begin{equation}\label{InteractionTerm}
H_{\text{int}}=\sum _{\bm{p},\bm{q}}\frac{1}{2} V\left(\bm{\kappa}\right)  a_{\bm{p}+\bm{\kappa}}^{\dagger} a_{\bm{q}-\bm{\kappa}}^{\dagger} a_{\bm{p}} a_{\bm{q}}.
\end{equation}
The zero-momentum term of interaction potential is
\begin{equation}
V_{0}=\frac{4\pi \hbar^{2} s}{m_{\gamma}},
\end{equation}
where $s=g/2\pi$ is the s-wave scattering length and $m_{\gamma}$ is effective mass of photons\cite{BEC03}.
As stated above, in BEC state, $N_{0}$ is relatively large, terms with $a^{\dag}_{0}a_{0}=N_{0}$ dominates. In Eq. \eqref{InteractionTerm}, only interaction terms with $N_{0}$ and $N_{0}^{2}$ are preserved, so the Hamiltonian $H$ can then be written as
\begin{equation}\label{InteractionSeparated}
\begin{aligned}
H  \thickapprox ~ & \epsilon'(0)  +\sum_{\bm{p}\neq0}\epsilon'(\bm{p})a_{\bm{p}}^{\dagger}a_{-\bm{p}}\\
 &   +\sum_{\bm{p}\neq0}N_{0}V_{\bm{p}}(a_{\bm{p}}^{\dagger}a_{-\bm{p}}^{\dagger}+a_{\bm{p}}a_{-\bm{p}}),
\end{aligned}
\end{equation}
where
\begin{equation}\label{Epr0}
\epsilon'(0)= N_0\epsilon(0)-\frac{1}{2}V_0 N_0^2
\end{equation}
and
\begin{equation}\label{epsilonprime}
\epsilon'(\bm{p}) =\epsilon(\bm{p})+N_{0}V_{\bm{p}} + N_{0} V_{0} - \mu,
\end{equation}
with $\epsilon'(\bm{p})$ being the modified photon energy.
\par Following Bogoliubov\cite{Bogo47}, we now introduce the following canonical transformation in order to diagonalize the quadratic-form Hamiltonian $H$ in Eq. \eqref{InteractionSeparated},
\begin{equation}\label{BogoTrans}
\begin{aligned}
b_{\bm{p}} &= u_{\bm{p}} a_{\bm{p}}+v_{\bm{p}} a_{-\bm{p}}^{\dagger },\\
b_{\bm{p}}^{\dagger} &= u_{\bm{p}} a_{\bm{p}}^{\dagger}+v_{\bm{p}} a_{-\bm{p}},
\end{aligned}
\end{equation}
where
$$u_{\bm{p}}^2-v_{\bm{p}}^2=1,$$
and $b_{\bm{p}}$, $b_{\bm{q}}^{\dagger}$ are respectively the annihilation and creation operators of the quasi-particles with momentum $\bm{p}$. They also obey the Bose commutation relations
\begin{equation}\label{BoseCMTRelation2}
\begin{aligned}
\left[b_{\bm{p}},b_{\bm{q}}^{\dagger}\right] & =  \delta_{\bm{p},\bm{q}},\\
\left[b_{\bm{p}},b_{\bm{q}}\right] & =  \left[b_{\bm{p}}^{\dagger},b_{\bm{q}}^{\dagger}\right] =  0.
\end{aligned}
\end{equation}
Then the diagonalized form of the Hamiltonian is given by
\begin{equation}\label{Hi2}
H=\sum _{\bm{p}} \tilde{\epsilon} _{\bm{p}} \left(b_{\bm{p}} b_{\bm{p}}^{\dagger }+\frac{1}{2}\right)+ E_{0}
\end{equation}
where $E_{0}$ is the ground-state energy of quasi-particles and $\tilde{\epsilon} _{\bm{p}}$ is the energy of quasi-particle with momentum $\bm{p}$. The equation of motion for the operators of the quasi-particles in Heisenberg picture is given by
\begin{equation}\label{Motion}
\begin{aligned}i\hbar\frac{\text{d}b_{\bm{p}}}{\text{d}t}  =  \left[b_{\bm{p}},H\right]
  =  \tilde{\epsilon}(\bm{p})b_{\bm{p}}.\end{aligned}
\end{equation}
By substituting Eqs. \eqref{BogoTrans} and \eqref{BoseCMTRelation2} into the original Hamiltonian in Eq. \eqref{InteractionSeparated} and comparing with Eq. \eqref{Hi2}, we get the following diagonalization conditions
\begin{equation}
\begin{aligned}
u_{\bm{p}}v_{\bm{p}} & = \frac{1}{2}N_{0} V_{\bm{p}}/\tilde{\epsilon}(\bm{p}),\\
u_{\bm{p}}^{2} & = \frac{1}{2}[\epsilon'(\bm{p})/\tilde{\epsilon}(\bm{p})+1],\\
v_{\bm{p}}^{2} & = \frac{1}{2}[\epsilon'(\bm{p})/\tilde{\epsilon}(\bm{p})-1].
\end{aligned}
\end{equation}
By solving Eq. \eqref{epsilonprime} and Eq. \eqref{Motion} for $\tilde{\epsilon}(\bm{p})$, we have\\
\begin{equation}\label{epsilon2}
\begin{aligned}
\tilde{\epsilon}(\bm{p})^{2} & =  \epsilon'(\bm{p})^{2}-N_{0}^{2}V_{0}^{2}\\
 & =  \epsilon(\bm{p})^{2}+2\epsilon(\bm{p})N_{0}V_{0},
\end{aligned}
\end{equation}
\begin{equation}\label{epsilon}
\tilde{\epsilon}(\bm{p})=\sqrt{\frac{\bm{p}^{2}N_{0}V_{0}}{m_{\gamma}}+\frac{\bm{p}^{4}}{4m_{\gamma}^{2}}}.
\end{equation}
Eq. \eqref{epsilon} is the Bogoliubov dispersion relation for the present system. Specifically, at critical temperature, the two terms under the square-root sign in Eq. \eqref{epsilon} are equal, so that the critical momentum $\bm{p}_{c}=2\sqrt{m_{\gamma}N_{0}V_{\bm{p}_{c}}}$ \cite{Chiao00}. At low temperature, we adopt the long-wavelength approximation
\begin{equation}\label{epsilonk}
\tilde{\epsilon}_{\bm{p}}=\tilde{c} \left|\bm{p}\right|
\end{equation}
where $\tilde{c} = \sqrt{N_0V_0/m_{\gamma}}$ is the sound velocity\cite{BEC03}. Actually, the fluctuation of photon gas described by $\tilde{c}$ and $b_{\bm{p}},~b_{\bm{p}}^{\dag}$ suggests that the excitations is phonon. From Ref. \cite{BEC03}, $\tilde{c}$ can be reexpressed as
\begin{equation}
\tilde{c}=\sqrt{\frac{V_0}{m_{\gamma}}}\sqrt{N-\alpha T}
\end{equation}
where $N$ is the total number of photons in cavity and $\alpha = m_{\gamma} k_{\text{B}} \sum_{\bm{p} \neq0} \bm{p}^{-2}$ is a constant independent of $\bm{p}$ and temperature $T$.
\subsection{Condensation Condition}
We now examine the conditions of BEC. All phase transitions have critical points, here is the critical temperature $T_{c}$. Only at low temperature, significant fraction of particles occupies the ground state. Above $T_{c}$, the condensate phase vanishes and the photons are free. For a normal three-dimensional BEc, the critical temperature is \cite{kittel80}
%To better illustrate the physical interpretation of modification factor $F_{\text{BEC}}$, we notice that apart from temperature it only concerns the effective mass of photon $m_{\gamma}$ and the configuration of microcavity, the total number of particles $N$, constant $\alpha$ that concerns the cut-off frequency $\omega_{\text{cut-off}}$ and the zero momentum potential $V_{0}$. In a non-relativistic case, energy shift demonstrates a positive correlation with $T$. The temperature-rise period has a upper limit $T_c$ that BEc disintegrates and all photons are free above it. The critical temperature $T_c$ is\cite{kittel80}
\begin{equation}
T_c = \frac{3.31\hbar^2}{\tilde{g}^\frac{2}{3}m_{\gamma} k_{\text{B}}}\frac{N}{\mathcal{V}}^{\frac{2}{3}}
\end{equation}
where $\mathcal{V}$ is the volume of the resonator.
However, in two-dimensional harmonic tarp, a revising factor $\sqrt{ \frac{6}{\pi^2}}\approx0.78$ is applied by using hard core calculations, see Eq. (18) of Ref. \cite{mli97}. For photons, the degeneracy $\tilde{g}=2$. Calculation shows that in the present configuration, the critical temperature is about $578.062 \text{~K}$ (for data see the caption of FIG.2). That indicates experiment can be performed even in room temperature. However, there exists another condition that the quantum concentration $\tilde{n}$ should satisfy
\begin{equation}
\begin{aligned}
\tilde{n}>\tilde{n}_{Q} & = & (\frac{m_{\gamma}k_{\text{B}}T}{2\pi\hbar^2})^{3/2},\end{aligned}
\end{equation}
where 
\begin{equation}
\begin{aligned}
\tilde{n} & = & N_{0}(\frac{2\pi\hbar^{2}}{m_{\gamma}k_{\text{B}}T})^{-3/2}
\end{aligned}
\end{equation}
and $\tilde{n}_{Q}$ is the critical quantum concentration.
In the present configuration, $\tilde{n}_Q=1.00828\times10^{15}~\text{m}^{-3}$ and $\tilde{n}=7.76374\times10^{18}~\text{m}^{-3}$. The condition is also satisfied.
\par Here, the dye solution serves as the heat bath and equilibrates the transverse modal degrees of freedom of photons and dye molecules via absorption and re-emission processes. At room temperature, photon frequency is above the low-frequency cut-off and can not be altered by temperature of the dye solution. In contrast to the case of a normal blackbody radiator, the photon number is determined by Stefan-Boltzmann law thus reveals a temperature dependency \cite{klaers10}.
\section{Dynamic Stark Shift}
Now, we consider an atom immersed in the dye solution, as shown in FIG. 1. The Hamiltonian of the system containing atom and photons is given by
\begin{equation}\label{HamiltonianAll}
H_{\text{sys}}=H_{\text{free}}+H+H'_{\text{int}},
\end{equation}
where $H_{\text{free}}$ is the Hamiltonian of a bare atom, $H$ is the Hamiltonian of the photonic system as given in Eq. \eqref{Hi2}, and $H'_{\text{int}}$ is the interaction Hamiltonian between atom and electric field. The atom can either be at high-lying (Rydberg) state or low-lying state. Typically, we consider an atom with only one electron in its outermost electron shell. Atoms like hydrogen and alkali like rubidium, potassium are widely used in experiments. Much work on high-lying state atom has already been done by Hollberg \textit{et al.} \cite{Hol84} and Zimmerman \textit{et al.} \cite{Zim79} both theoretically and experimentally. Here, we only consider the case of a high-lying atom. The unperturbed atomic Hamiltonian obeys the eigenvalue equation
\begin{equation}
H_{\text{free}}\left|m\right\rangle = E_m \left|m\right\rangle.
\end{equation}
Our interest now is focused on $H'_{\text{int}}$, the interaction Hamiltonian between atom and photon fluid is
\begin{equation}\label{hpint}
H'_{\text{int}}=-e \bm{r} \cdot \bm{E},
\end{equation}
where $\bm{r}$ is the radius vector of the electron, $\bm{E}$ is the electric field induced by BEc of photons, and $e$ is the absolute value of the electron charge.
To get the expression of $H'_{\text{int}}$, we need the expression of $\bm{E}$. Generally, a vector potential of the electromagnetic field is
\begin{equation}\label{GeneralVectorPotential}
\bm{A}=\sum_{\bm{p}}\sqrt{\frac{\hbar}{2\mathcal{V}\varepsilon_{0}\omega_{\bm{p}}}}\left(a_{\bm{p}}(t)+a_{\bm{p}}^{\dagger}(t)\right)\hat{e}_{\bm{p}}.
\end{equation}
It is worth noting that we add ``$(t)$'' to emphasize that the operators $a_{\bm{p}}$ has a time-dependent term $e^{-i\omega_{\bm{p}}t}$. For $b_{\bm{p}}$, it is $e^{-i\tilde{\omega}_{\bm{p}}t}$ accordingly. $\tilde{\omega}_{\bm{p}}$ is the angular frequency of phonon with momentum $\bm{p}$.
By using Bogoliubov transform (Eq. \eqref{BogoTrans}), we can rearrange the vector potential of the photons and it is given by
\begin{equation}\label{ETrans}
\bm{A}=\sum_{\bm{p}}\sqrt{\frac{\hbar}{2\mathcal{V}\varepsilon_{0}\omega_{\bm{p}}}}e^{\beta}\left(b_{\bm{p}}(t)+b_{\bm{p}}^{\dagger}(t)\right)\hat{e}_{\bm{p}}
\end{equation}
where
\begin{equation}\label{ebeta}
e^{\beta} = u_{\bm{p}} + v_{\bm{p}}=\sqrt[4]{1+\frac{N_{0} V_{\bm{p}}}{\epsilon_{\bm{p}}}}.
\end{equation}
At critical momentum $\bm{p}_c$, the interaction potential has a harmonic form\cite{Chiao00}
\begin{equation}\label{vpc}
\begin{aligned}
V_{\bm{p}}=\lambda \bm{p}^2,
\end{aligned}
\end{equation}
where $\lambda = 1/4 m_{\gamma} N_{0}$ is a constant with a unit of mass$^{-1}$. As a reasonable assumption, we substitute the expression of $V_{\bm{p}}$ in Eq. \eqref{vpc} into Eq. \eqref{ebeta} and get
\begin{equation}
\begin{aligned}
e^{2\beta} & =  \sqrt{1+\frac{2N_{0}V_{\bm{p}}}{\epsilon_{\bm{p}}}}\\
 & =  \sqrt{1+\frac{2N_{0}\lambda\bm{p}^{2}}{\bm{p}^{2}/2m_{\gamma}}}\\
 & =  \sqrt{1+4m_{\gamma}N_{0}\lambda}\\
 & =  \sqrt{2}.\end{aligned}.
\end{equation}
By using the relation $\bm{E}=-\frac{\partial \bm{A}}{\partial t}$, the electric field in Eq. \eqref{hpint} can be obtained
\begin{equation}\label{BECElectricPotential}
\bm{E}=i\sum_{\bm{p}}\sqrt{\frac{\hbar}{2\mathcal{V}\varepsilon_{0}\omega_{\bm{p}}}}e^{\beta}\tilde{\omega}_{\bm{p}}\left(b_{\bm{p}}(t)-b_{\bm{p}}^{\dagger}(t)\right)\hat{e}_{\bm{p}}.
\end{equation}
Subsequently, the interaction Hamiltonian is given by
\begin{equation}\label{intH}
\begin{aligned}
H'_{\text{int}}=-i e\sum_{\bm{p}}\sqrt{\frac{\hbar}{2\mathcal{V}\varepsilon_{0}\omega_{\bm{p}}}}e^{\beta}\tilde{\omega}_{\bm{p}}(b_{\bm{p}}(t)-b_{\bm{p}}^{\dagger}(t))\bm{r}\cdot\hat{e}_{\bm{p}}.
\end{aligned}
\end{equation}
\par Since external electromagnetic field is weak compared with the internal atomic interactions, one may use the perturbation theory. From the second-order perturbation theory, the energy shift for a reference state $|\phi,\psi_{\bm{p}}^{n}\rangle$ is given by
\begin{equation}\label{stark}
\begin{aligned}\Delta E(\phi) & =\begin{aligned}\sum_{n',I,\bm{p}} & \frac{\left|\left\langle \phi,\psi_{\bm{p}}^{n}\left|H'_{int}\right|I,\psi_{\bm{p}}^{n'}\right\rangle \right|{}^{2}}{E^{\psi_{\bm{p}}^{n}}_{\phi}-E^{\psi_{\bm{p}}^{n'}}_{I}}\end{aligned}
\\
 & \begin{aligned}=\sum_{I,\bm{p}}\Big( & \frac{\left|\left\langle \phi,\psi_{\bm{p}}^{n}\left|H'_{int}\right|I,\psi_{\bm{p}}^{n-1}\right\rangle \right|{}^{2} N_{\bm{p}}}{E_{\phi}+N_{\bm{p}}\epsilon_{\bm{p}}'-\left(E_{I}+(N_{\bm{p}}-1)\epsilon_{\bm{p}}'\right)}+\\
 & \frac{\left|\left\langle \phi,\psi_{\bm{p}}^{n}\left|H'_{int}\right|I,\psi_{\bm{p}}^{n+1}\right\rangle \right|{}^{2}}{E_{\phi}+N_{\bm{p}}\epsilon_{\bm{p}}'-\left(E_{I}+(N_{\bm{p}}+1)\epsilon_{\bm{p}}'\right)}\Big)\end{aligned}
\\
 & \begin{aligned}=\sum_{I,\bm{p}}e^{2\beta}\frac{\hbar}{2\mathcal{V}\varepsilon_{0}\omega_{\bm{p}}}\tilde{\omega}_{\bm{p}}^{2}\Big( & \frac{\left|\left\langle \phi\left|\bm{r}\cdot\hat{e}_{\bm{p}}\right|I\right\rangle \right|{}^{2}}{E_{\phi}-E_{I}\pm\epsilon_{\bm{p}}'}N_{p}\\
+\frac{\left|\left\langle \phi\left|\bm{r}\cdot\hat{e}_{\bm{p}}\right|I\right\rangle \right|{}^{2}}{E_{\phi}-E_{I}-\epsilon_{\bm{p}}'}\Big)
\end{aligned}
\\
 & =\Delta E_{\phi}^{\text{AC}}+\Delta E_{\phi}^{\text{Lamb}},\end{aligned}
\end{equation}
where ${N}_{p} = 1/(e^{\hbar \tilde{\omega}/k_{B}T}-1)$. It is worth noting that in result Eq. \eqref{stark}, two terms are presented.
The first term is the dynamic (or AC) Stark shift induced by real photons and the second one is the Lamb shift induced by virtual photons, or by vacuum.
The Lamb shift $\Delta E_{\phi}^{\text{Lamb}}$ includes both measurable part and the mass renormalization term \cite{bethe47}. After changing the summation over the momentum $\bm{p}$ to a two-dimensional integral, we have
\begin{equation}\label{DEAC}
\begin{aligned}\Delta  E^{\text{AC}} & =  \sqrt{2}\frac{e^{2}Kc}{4\pi\varepsilon_{0}\tilde{c}}(\frac{k_{\text{B}}T}{\hbar c})^{2}\sum_{I}\text{P}\int\text{d}t\left|\bm{r}_{\phi,I}\right|^{2}\frac{t^{2}}{e^{t}-1}\times\\
 & \Big(\frac{1}{(E_{\phi}-E_{I})/k_{\text{B}}T+t}+\frac{1}{(E_{\phi}-E_{I})/k_{\text{B}}T-t}\Big)\\
\end{aligned}
\end{equation}
where $t = p\tilde{c}/k_{\text{B}}T$, $\left|\bm{r}_{\phi,I}\right|^{2}=\left|\left\langle \phi\left|\bm{r}\right|I\right\rangle\right|^{2} $, $K=\int k_{z}(r)\text{d}r$ and
\begin{equation}\label{DELamb}
\begin{aligned}\Delta E^{\text{Lamb}}= & \sqrt{2}\frac{e^{2}Kc}{4\pi\varepsilon_{0}\tilde{c}}(\frac{k_{\text{B}}T}{\hbar c})^{2}\sum_{I}\text{P}\int\text{d}t\times\\
 & \left|\bm{r}_{\phi,I}\right|^{2}\frac{t^{2}}{(E_{\phi}-E_{I})/k_{\text{B}}T-t}.
\end{aligned}
\end{equation}
``P'' in the above equations denotes that a principle value integral should be preformed, since it is an improper integral.
\par By using the vector potential in a normal fluid (Eq.\eqref{GeneralVectorPotential}) and its corresponding electric field, we can easily get the dynamic Stark shift in a normal two-dimensional photon gas
\begin{equation}\label{DEACNormal}
\begin{aligned}\Delta E_{\text{Normal}}^{\text{AC}}&=  \frac{e^{2}K}{4\pi\varepsilon_{0}}(\frac{k_{\text{B}}T}{\hbar c})^{2}\sum_{I}\text{P}\int\text{d}t\left|\bm{r}_{\phi,I}\right|^{2}\frac{t^{2}}{e^{t}-1}\times\\
 & \Big(\frac{1}{(E_{\phi}-E_{I})/k_{\text{B}}T+t}+\frac{1}{(E_{\phi}-E_{I})/k_{\text{B}}T-t}\Big).\end{aligned}
\end{equation}
Comparing Eq. \eqref{DEAC} with Eq. \eqref{DEACNormal}, we can rearrange the shift in form of
\begin{equation}\label{F-defractorization}
\begin{aligned}
\Delta E^{\text{AC}}=F_{\text{BEC}}\cdot \Delta E_{\text{Normal}}^{\text{AC}}
\end{aligned}
\end{equation}
where
\begin{equation}\label{F}
\begin{aligned}F_{\text{BEC}} & =\sqrt{2}\cdot\frac{c}{\tilde{c}}\end{aligned}
\end{equation}
is the modification factor in photonic BEc. The factor $F_{\text{BEC}}$ describs the modification with respect to the Stark shift in a normal photon fluid. \\
\section{Results and discussions}
From Eq. \eqref{F-defractorization}, we can see that the Stark shift in a system of BEC of photons is modified by a factor $F_{\text{BEC}}$, compared to that in a normal two-dimensional photon fluid.
To better illustrate the physical interpretation, we can rewrite it in the following form,
\begin{equation}
F_{\text{BEC}}=\sqrt{\frac{2m_{\gamma}}{V_{0}}}\frac{c}{\sqrt{N-\alpha T}}.
\end{equation}
\par In order to give a numerical impression of $F_{\text{BEC}}$, we take the experimental data from Klaers \textit{et al.} \cite{klaers10}. The effective mass $m_{\gamma}\thickapprox 6.7\times 10^{-36}~\text{kg}$, the zero momentum interaction potential $V_{0}\thickapprox 3.2\times 10^{-33}~\text{J}$, the speed of light $c\thickapprox 3\times 10^{8}~\text{m/s}$, and the total photon number $N$ no less than $77,000$. As a result, $F_{\text{BEC}}$ is a number of order $4\sim5$.
\begin{figure}[htbp]
\small
\centering
\includegraphics[width=3.4 in]{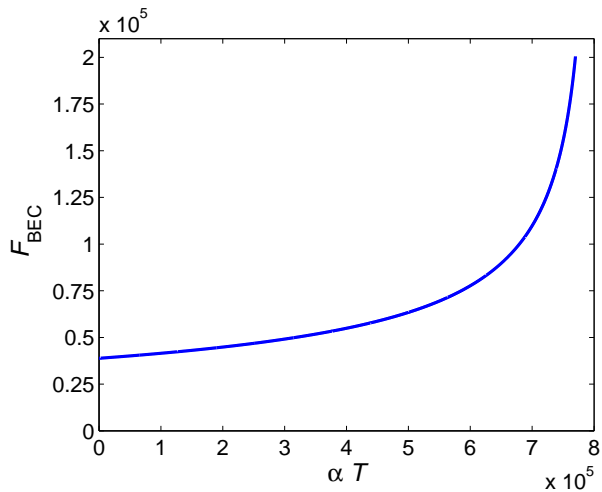}
\caption{(Color online) Modification factor $F_{\text{BEC}}$ as a function of modified temperature $\alpha T$. Data are based on the experiment performed by Klaers \emph{et al.}\cite{klaers10}. $V_{0}=4\pi \hbar^{2} s/m_{\gamma}\thickapprox 3.2\times 10^{-33}~\text{J}$, cut-off frequency $\omega_{\text{cut-off}}\thickapprox 3\times 10^{15}~\text{Hz}$, effective mass of photon $m_{\gamma}\thickapprox6.7\times 10^{-36}~\text{kg}$, dimensionless parameter for interaction strength $g\thickapprox10^{-3}$, and the volume $\mathcal{V} = 1.45\times 10^{-12}~\text{m}^{3}$.}
\label{fig:F-T}
\end{figure}
FIG. 2 shows the function of $F_{\text{BEC}}$ to $\alpha T$. At near absolute zero temperature, the modification factor $F_{\text{BEC}}$ does not vanish but approximates $7.6574\times10^{4}$. In spite that the normal shift is small, the total effect might be significant. So, as long as the fluid is in BEC state, the shift is much larger than the normal one.
\par Up to now, one may raise the question why the dynamic Stark shift is so large in BEC of photons. It can be answered as follows. As is known, the dynamic Stark shift is caused by the perturbation of electric field. The key physical function concerning the atomic QED is the density of state (DOS) of the photonic system. In a non-absorbing linear medium, the DOS is given by
\begin{equation}
\begin{aligned}
\rho(\omega_{\bm{p}}) = \mathcal{V} \omega_{\bm{p}}^{2} / \pi^{2} c^{3}.
\end{aligned}
\end{equation}
However, in the present system, the photons are replaced by quasi-particles. The DOS of photons is changed accordingly to that of quasi-particles
\begin{equation}
\begin{aligned}
\rho(\tilde{\omega}_{\bm{p}}) = \mathcal{V} \tilde{\omega}_{\bm{p}}^{2} / \pi^{2} \tilde{c}^{3}.
\end{aligned}
\end{equation}
We can see that the DOS in BEC of photons is a function of the velocity of quasi-particles $\tilde{c}$ which is a temperature dependent value. At near zero temperature, $\tilde{c}$ is much smaller than $c$, thus the DOS can be relatively much larger than that in a normal fluid. Much concentrated quasi-particles results in a heavily enhanced Stark shift.\par
%-----------------------------------HERE NEEDS THE EXPLANATION ON SQUEEZING EFFECT------------------------------
%This inspired us on its potential application to compensate the Doppler shift in a variety of atom trap system. Since the Doppler shift is velocity dependent, it can be large enough to overwhelm the range of shift induced by Zeeman effect which is often seen in compensating relatively small Doppler shift. More discussions on this topic can be found at the last section of this paper.
%-----------THIS PART NEEDS MORE CITES ON SHIFT-COMPENSATING TOPICS, SEE WANGYIQIU's WORK!------------------
Additionally, we consider the role of dimension of the electromagnetic environment on the dynamic Stark shift. Unlike usual photon systems, photon fluid in the present system has a non-vanishing chemical potential, which is significantly different from the three-dimensional photon system typically encountered in Planck problem. This may owe to the repulsive pairwise interaction between photons in a BEc. To find the chemical potential $\mu$, we use
\begin{equation}
\mu=\frac{\partial \epsilon_0}{\partial N},
\end{equation}
where $\epsilon_0$ is the ground-state energy $\left\langle \psi_0 \left|H\right|\psi_0 \right\rangle $. Since it is in BEC state, the zero-momentum terms dominates, thus the other terms can be dropped out. In Eq. \eqref{Epr0}, we have $\epsilon_0=\frac{1}{2}N^2V_0$ and can then get
\begin{equation}\label{chempot}
\mu=N V_0 \approx N_0 V_0
\end{equation}
Eq. \eqref{chempot} clearly shows that a non-vanishing chemical potential exists in this resonator.
The existence of low-dimensional BEC has already been discussed. Back to 1997, Mullin \cite{mli97} theoretically showed that the Bose gas in a two-dimensional harmonic trap in the thermodynamic limit shows a phase transition at some critical temperature $T_{\text{c}}$. What is more, he pointed out that for dimensions no less than 1, $T_{\text{c}}$ has the same form of expression. However, in one-dimensional case, there is no condensation.\par
The dynamic Stark shift in a three-dimensional normal photon fluid is given by \cite{farley81}
\begin{equation}\label{farley}
\begin{aligned}
\frac{e^{2}}{6\pi^{2}\varepsilon_{0}}(\frac{k_{\text{B}}T}{\hbar c})^{3}\sum_{I} & \text{P}\int\text{d}t\frac{t^{3}}{e^{t}-1}\Big(\frac{1}{(E_{\phi}-E_{I})/k_{\text{B}}T+t}\\
 & +\frac{1}{(E_{\phi}-E_{I})/k_{\text{B}}T-t}\Big).\end{aligned}
\end{equation}
Comparing the Stark shift in Eq. \eqref{farley} with that in a two-dimensional case in Eq. \eqref{DEACNormal}, we can see that for Stark shift in systems of different dimensions, the dimensionality affects the value of the shift.
\par
In recent years, the technology of atom manipulation by laser has been developing rapidly. Stark shift is considered to be suitable to compensate the Doppler shift of traveling atom. In a normal electromagnetic environment, the Stark shift is too small to be qualified. However, in the BEC of photons, the heavily enhanced Stark shift seems promising. Due to its large value, a wide range of frequency in Doppler compensation might be possible. What is more, the heavily enhanced Stark shift might be used for tunable far-infrared photodetectors \cite{Huang95}.
\section{Conclusion}
In summary, we have investigated the dynamic Stark shift of a high-lying atom in a system of BEC of photons within the framework of nonrelativistic QED theory. The effective mass and the chemical potential of photons inside a two-dimensional microcavity are nonvanishing. By using Bogoliubov transform, sound waves exists for long-wavelength disturbances of the system. It is found that compared to that in two-dimensional free space, Stark shift in photonic BEC is modified by a factor $F_{\text{BEC}}$, which is a monotonically increasing function of temperature $T$ and depends on a few other parameters of the system. Below critical temperature $T_{c}$, the value of Stark shift is always greater than that in a normal two-dimensional photonic fluid ($4\sim5$ orders of magnitude). Physical origin of this phenomenon is discussed and potential applications are also proposed including Doppler shift compensation and tunable far-infrared photodetectors. It is hoped that the predicted properties will be verified in physics laboratories for the not too distant future.
\begin{acknowledgments}
This work was supported by the National Natural Science Foundation of China(No.11204088), Fundamental Research Funds for the Central Universities(No. 2011ZM0091),
and SRP of South China University of Technology.
\end{acknowledgments}

\bibliography{Bibliography}% Produces the bibliography via BibTeX.

\end{document}